\documentclass[11pt]{article}  
  
\usepackage{amssymb}  
\usepackage{amsmath}  
\usepackage{amsthm}  
\usepackage{latexsym}  
  
    \newcommand{\Reals}{\it I\kern-.4emR}  
      
    \newcommand{\Notin}{/\kern-.6em\hbox{$\in$}}  
    \newcommand{\Notequiv}{/\kern-.6em\hbox{$\equiv$}}  
      
    \newcommand{\Ceals}{\it I\kern-.65emC}  
    \newcommand{\MM}{\it I\kern-.4emM}  
    \newcommand{\NN}{\it I\kern-.4emN}  
    \newcommand{\yy}{\it Y\kern-.8emY}  
    \newcommand{\zz}{\makebox[.80em]{\it Z\kern-.46emZ}}  
    \newcommand{\tzz}{\makebox[.80em]{\scriptsize\it Z\kern-.46emZ}}  
  
\newcommand{\R}[0]{{\mathbb{R}}}  
  
\def\C{{\mathbb{C}}}  
  
\def\R{{\mathbb{R}}}

\def\K{{\mathbb{K}}}

\newcommand{\Aut}[0]{{\rm Aut}}

\newcommand{\cB}[0]{{\mathcal B}}  
\newcommand{\cL}[0]{{\mathcal L}}  
\newcommand{\cK}[0]{{\mathcal L}}

\newcommand{\cC}[0]{{\mathcal C}}

\newcommand{\cH}{{\cal H}}  
  
\newcommand{\cD}{{\cal D}}

\newcommand{\cP}{{\cal P}}

\newcommand{\onemat}[0]{{\mathbf 1}}  
  
\newcommand{\tr}[0]{{\rm Tr}}  
\newcommand{\ad}[0]{{\rm ad}}  
\newcommand{\ket}[1]{|#1\rangle}  
\newcommand{\bra}[1]{\langle #1|}  
\newcommand{\scalar}[2]{\langle #1|#2\rangle}

    \newtheorem{theorem}{Theorem}[section]  
    \newtheorem{lemma}[theorem]{Lemma}  
      
    \newtheorem{proposition}[theorem]{Proposition}  
      
    \newtheorem{example}[theorem]{Example}  
      
    \newtheorem{observation}[theorem]{Observation}  
    \newtheorem{conjecture}[theorem]{Conjecture}  
    \newtheorem{definition}[theorem]{Definition}  
\title{\Large \textbf{Mutually Unbiased Bases and Orthogonal  
Decompositions of Lie algebras}}   
  
\author{
P. Oscar Boykin$^1$,
Meera Sitharam$^2$,
Pham Huu Tiep$^3$,
Pawel Wocjan$^4$\footnote{corresponding author wocjan@cs.caltech.edu}\\
{\small 1. Dept. of Electrical and Computer Engineering,
           University of Florida,
           Gainesville, FL 32611}\\
{\small 2. Computer and Information Science and Engineering,
           University of Florida,
           Gainesville, FL 32611}\\
{\small 3. Department of Mathematics,
           University of Florida,
           Gainesville, FL 32611}\\
{\small 4. Institute for Quantum Information,
           California Institute of Technology,
           Pasadena, CA}\\
}
  
\begin{document}  
  
\maketitle  
\thispagestyle{empty}

\begin{abstract}  
We establish a connection between the problem of constructing maximal  
collections of mutually unbiased bases (MUBs)  and an open  
problem in the theory of Lie algebras. More precisely, we show that a  
%---------------1  
collection of $\mu$ MUBs in $\K^n$ gives rise to a collection of $\mu$  
Cartan subalgebras of the special linear Lie algebra $sl_n(\K)$ that  
are pairwise orthogonal with respect to the Killing form, where  
$\K=\R$ or $\K=\C$. In particular, a complete collection of MUBs in $\C^n$  
gives rise to a so-called orthogonal decomposition (OD) of $sl_n(\C)$.   
%---------------2  
The converse holds if the Cartan subalgebras in the OD are also $\dag$-closed,  
i.e., closed under the adjoint operation.   
%---------------3  
In this case, the Cartan subalgebras  
have unitary bases, and the above correspondence becomes equivalent  
to a result of \cite{bbrv02} relating collections of MUBs to   
collections of maximal commuting classes of unitary error bases, i.e.,   
orthogonal unitary matrices.  
  
It is a longstanding conjecture that ODs of $sl_n(\C)$ can only  
exist if $n$ is a prime power. This corroborates further the general  
belief that a complete collection of MUBs can only exist in prime power  
dimensions. The connection to ODs of $sl_n(\C)$  
potentially allows the application of known results on (partial)  
ODs of $sl_n(\C)$ to MUBs.   
  
%---------------4  
As a first example,   
a known result on ODs shows that for the first non-prime-power   
dimension $n=6$, at most three MUBs   
can be obtained from maximal commuting classes of monomial matrices,  
%---------------5  
which are used in   
all known constructions of MUBs.  
  
%----------------6  
As a second example, it is known that so-called irreducible ODs exist    
only for prime power dimensions   
and are essentially unique (except for  
$n = 27$), namely, a standard or canonical OD
that is based on monomial matrices. This OD is additionally $\dag$-closed,   
corresponds to a partition of a ``nice'' \cite{ACW:04} unitary error basis   
%-----------------7  
and is essentially equivalent to all the known partitions including those   
of \cite{bbrv02}, which yield the known complete collections of MUBs  
for prime power dimensions.  
%-----------------8  
A corollary of these results is that for each dimension $n \le 5$, an   
essentially unique, complete collection of MUBs exists.   
  
Intuitively, a complete collection of MUBs that   
corresponds to an irreducible OD possesses a large group of symmetries.   
Formally, by irreducible, we mean the   
subgroup of $\Aut(sl_n(\C))$ - of automorphisms that additionally preserve  
the OD - acts on $sl_n(\C)$ irreducibly;  and   
uniqueness or equivalence are under the action of   
$\Aut(sl_n(\C))$, i.e.,   up to    
$\Aut(sl_n(\C))$-conjugacy.  
\end{abstract}

  
%%% The main text  
  
%%%%%%%%%%%%%%%%%%%%%%%%%%%%%%%%%%%%%%%%%%%%%%%%%%%%%%%%%%%%  
%  
% Section: Introduction  
%  
%%%%%%%%%%%%%%%%%%%%%%%%%%%%%%%%%%%%%%%%%%%%%%%%%%%%%%%%%%%%  
  
\section{Introduction}  
\label{intro}
Two orthonormal bases ${\cal B}$ and ${\cal B}'$ of the Hilbert space  
$\C^n$ are called {\it mutually unbiased} if and only if  
\begin{equation}  
|\scalar{\phi}{\psi}|=1/\sqrt{n}  
\end{equation}  
for all $\ket{\phi}\in {\cal B}$ and all $\ket{\psi}\in {\cal  
B}'$.   
  
%This is lifted directly from the real MUB paper: 
The problem of determining bounds on the maximum number $M_{\C^d}$ of
bases over $\C^d$ that are mutually unbiased is an important
open problem\cite{KW:05} which has received much attention\cite{Ivanovic:81,
WF:89,bbrv02,KR:03,GHW:04, Wocjan}.
We
refer the reader to e.g.\ \cite{ACW:04} for an overview of known
bounds.

The most common application of MUBs is found in quantum
cryptography where MUBs are the quantum states used
in most QKD protocols\cite{bruss98,BP:00,CBKG:02,br04i}.
 
By putting the vectors of the bases ${\cal B}$ and ${\cal B}'$ as
columns of matrices $M_{{\cal B}}$ and $M_{{\cal B}'}$, the above
condition says that $M_{{\cal B}}^\dag M_{{\cal B}'}$ should also be a
generalized Hadamard (scaled by $1/\sqrt{n}$). The problem of
determining the maximal number of bases that are mutually unbiased is
an open problem. It is known that $n+1$ is an upper bound on the
number of mutually unbiased bases in dimension $n$.
  
For prime power dimensions, several constructions\cite{WF:89,bbrv02,KR:03}
attain this bound.
One construction\cite{bbrv02}, uses so-called ``nice'' error bases;   
\cite{ACW:04} shows  
the severe ``reduce-to-minimum-prime-power'' limitation of   
MUB constructions that use such special bases   
- for non-prime-power dimensions.   
For these dimensions, while improved constructions   
give somewhat better lower bounds such as \cite{Wocjan}, they   
are not known to be tight.  
It is generally believed, however,   
that the upper bound of $n+1$ can only be attained for prime  
power dimensions $n$.   
  
We show that this belief is reflected in a conjecture  
on orthogonal decompositions of complex simple Lie algebras into  
Cartan subalgebras.  
Establishing this connection is shown to be useful in at least two ways.  
First, known results on orthogonal Cartan subalgebras of   
Lie algebras yield three new results about MUBs for small  
dimensions $n \le 6$.  
Second, we obtain an alternative viewpoint  and new properties for   
known constructions of MUBs \cite{bbrv02} in prime power dimensions,  
obtained by partitioning nice   
\cite{ACW:04} unitary error bases.

%%%XXXXXXXXXXXXXXXXXXXMotivation for studying Orthogonal Decompositions
%%%of Lie algebras XXXXXXXXXXXXXXXX
The notion of {\it orthogonal decompositions} of simple complex Lie algebras 
emerges from the pioneering work of J. G. Thompson \cite{T1, T2}, where 
such a decomposition of
the Lie algebra of type $E_{8}$ played a crucial role in his construction of a sporadic 
finite simple group, nowadays called Thompson's group. 
Orthogonal decompositions turn out to be interesting not only 
by their inner geometric structures, but also by their interconnections with other areas of
mathematics: Lie algebras and Lie groups, finite groups, combinatorics and finite geometries,
quadratic forms and integral lattices. Following Thompson's work, the original motivation 
for studying orthogonal decompositions was to obtain a ``Lie'' realization of certain finite 
simple groups, as well as to find new interesting (in particular, unimodular) 
Euclidean lattices. Systematic investigation 
of orthogonal decompositions has been undertaken by a number of mathematicians; for a 
comprehensive treatment of the topic the reader is referred to \cite{KT:94}. 

%The aim of these notes
%is to demonstrate a connection between orthogonal decompositions of Lie algebras and 
%yet another object in mathematics and quantum information processing -- the 
%notion of mutually unbiased bases.   
  
%-----------------------  
\subsubsection*{Organization}  
In Section \ref{oscarpawel}, we give the relevant background on   
MUB constructions, their  
properties and their limitations.    
Section \ref{sec:monodef} defines the concept of monomial MUBs and
we see that existing MUB constructions are monomial.
Lie algebra preliminaries  are laid out in Section \ref{liebasic}.  
Section \ref{connection} draws the   
connection between orthogonal Cartan subalgebras of Lie algebras,   
MUBs and maximal commuting classes of orthogonal unitary matrices.  
Section \ref{monomial} utilizes this connection and a known result  
about Cartan subalgebras consisting of monomial matrices   
to give a new upper bound on a common type of MUBs for dimension $n=6$.  
Section \ref{irreducibility} elaborates on another known result  
about so-called irreducible orthogonal decompositions   
into Cartan subalgebras, which yields a new   
uniqueness result about MUBs for dimensions $n \le 5$,   
and uncovers new properties of   
known constructions of complete MUB collections   
obtained by partitioning nice error bases.  
Finally, we discuss the potential for extending these results.

\section{Background on MUB Constructions and Nice Error Bases}  
\label{oscarpawel}  
  
In this section we recall that the problem of constructing mutually  
unbiased bases corresponds to partitioning unitary matrices that are  
orthogonal with respect to the trace inner product into certain  
commuting classes. The mutually unbiased bases correspond to the  
common eigenvectors of the commuting classes.
\iffalse
%This may only confuse at this stage
Later it will become  
clear that the unitary matrices of each commuting classes (without the  
identity matrix) form a basis for the Cartan subalgebras of the  
orthogonal decomposition of $sl_n(\C)$. The fact that all matrices are  
unitary implies that all Cartan subalgebras are $\dagger$-closed.  
\fi
  
A {\em unitary error basis} ${\cal E}$ is a basis of the vector space
of complex $n\times n$ matrices that is orthogonal with respect to the
trace inner product.  In other words, a set of unitary matrices ${\cal
E}:=\{U_1=\onemat,U_2,\ldots,U_{n^2}\}$ is a unitary error basis iff
\begin{equation}  
  \tr(U_k^\dag U_l) = n \, \delta_{k,l} \,,\quad k,l \in  
\{1,\ldots,n^2\} \,.  
\end{equation}  
Two constructions of unitary error bases are known. The first are  
{\it nice} error bases, a  
group-theoretic construction due to Knill \cite{Knill:96a};  
 
The second type of unitary error bases are  
shift-and-multiply bases, a combinatorial construction due to Werner  
\cite{Werner:00}. There exist nice error bases that are not  
equivalent to any shift-and-multiply basis, as well as  
shift-and-multiply bases that are wicked (not nice) \cite{KR:03b}.  
  
\begin{definition}[Nice error basis] 
Let $G$ be a group of order $n^2$ with identity element $e$. A set 
${\cal N}:=\{U_g\in U_n(\C) \, : g \in G\}$ is a nice error basis if 
\begin{enumerate} 
\item $U_e$ is the identity matrix, 
\item $Tr(U_g)=0$ for all $g\in G\setminus\{e\}$, and 
\item $U_g U_h = \omega(g,h) U_{gh}$ for all $g,h\in G$,  
\end{enumerate} 
where $\omega(g,h)$ are complex numbers of modulus one.
\end{definition} 
The group $G$ is a called the {\em index group} of the nice error
basis ${\cal N}$ because its elements enumerate the elements of ${\cal
N}$. It is easily see that a nice error basis is a unitary error
basis. Observe that $U_g^\dagger=\omega(g^{-1},g)^{-1}
U_{g^{-1}}$. Assume that $g,h$ are distinct elements of $G$, then
$g^{-1}h\neq e$ and consequently $\tr(U_g^\dagger U_h) =
\omega(g^{-1},g)^{-1} \omega(g^{-1},h) \tr(U_{g^{-1} h})=0$ by the
second property of a nice error basis.
 
Unitary error bases can be used to produce mutually unbiased bases  
using the following construction:  
 
\begin{lemma}[Construction of MUBs]  
\label{lemma:constr_MUB}  
Let ${\cal C}=\cC_1\cup\ldots\cup\cC_\mu$ with $\cC_k\cap  
\cC_l=\{\onemat\}$ for $k\neq l$ be a set of $\mu(n-1)+1$ unitary  
matrices that are mutually orthogonal with respect to the trace inner  
product. Furthermore, let each class $\cC_k$ of the partition of  
${\cal C}$ contain $n$ commuting matrices $U_{k,t}$, $0\le t\le n-1$,  
where $U_{k,0}:=\onemat$. For fixed $k$, let ${\cal B}_k$ contain the  
common eigenvectors $|\psi_i^k\rangle$ of the matrices $U_{k,j}$. Then  
the bases ${\cal B}_k$ form a set of $\mu$ mutually unbiased bases,  
i.e.,  
\begin{equation}  
  |\langle\psi_i^k|\psi_j^l\rangle|^2 = 1/n \quad \mbox{ for $k\ne l$.}  
\label{eq:unbiased}  
\end{equation}  
The converse direction is also true. If there are $\mu$ mutually  
unbiased bases, then there are $\mu$ commuting classes with the above  
properties.  
\end{lemma}  
\noindent  
For a proof of this result, see \cite{bbrv02}.   
  
Observe that the commuting classes in the above construction are  
maximal. This is because there can be at most $n$ mutually commuting  
unitary matrices acting on $\C^n$ that are orthogonal with respect to  
the trace inner product. Let $\cC$ be a set of mutually commuting  
matrices. Since the matrices in $\cC$ are mutually commuting, they can  
be diagonalized simultaneously. The trace orthogonality of a unitary  
error basis implies that the diagonals of the elements of $\cC$, when  
written in their common eigenbasis, must be pairwise orthogonal as  
vectors in $\C^n$ with respect to the standard inner product. Since  
there can be at most $n$ orthogonal vectors, the above commuting  
classes are maximal.
  
It was shown in \cite{bbrv02} and in \cite{ACW:04} that unitary error
bases consisting of tensor products of generalized Pauli matrices (a
particular class of nice error bases) can be partitioned according to
Lemma~\ref{lemma:constr_MUB} so that we obtain a collection of $n+1$
MUBs for dimensions $n$ that are prime powers. Basically, the same
construction was already used in \cite{KT:94} for orthogonal
decompositions of $sl_n(\C)$.
  
Let $n=p^e$ be a prime power. Define the generalized Pauli operators 
acting on $\C^p$ 
\begin{equation}  
X := \sum_{k=0}^{p-1} |k\rangle\langle k+1|\,,   
\quad   
Z := \sum_{k=0}^{p-1} \omega^{k} |k\rangle\langle k|\,,   
\end{equation}  
where $\omega$ is a $p$th root of unity. For $(x,y):=(x_1,\ldots,x_e,  
z_1,\ldots,z_e)\in Z_p^e\times Z_p^e$ define the tensor product  
$U^{(x,y)}$ of generalized Pauli matrices to be   
\begin{equation}  
U^{(x,y)} := X^{x_1} Z^{z_1}\otimes \cdots \otimes X^{x_e} Z^{z_e} \,,  
\label{eq:tensorGenPauli}  
\end{equation}  
where $Z_p:=\{0,\ldots,p-1\}$ is the cyclic group of order $p$ and 
$Z_p^e$ is the direct product of $e$ copies of $Z_p$. Then it is 
readily verified that the set ${\cal N}:=\{ U^{(x,z)} : (x,z)\in 
Z_p^e\times Z_p^e\}$ is a unitary error basis for $\C^n$. Moreover, it 
is a nice error basis with index group $G:=Z_p^e\times Z_p^e$. 
  
\begin{theorem}[Complete sets of MUBs]\label{th:oscarcomplete}  
Let $n=p^e$ be a prime power dimension. Then the nice error basis 
in eq.~(\ref{eq:tensorGenPauli}) consisting of tensor products of 
generalized Pauli matrices can be partitioned according to 
Lemma~\ref{lemma:constr_MUB} into $n+1$ commuting classes showing that 
there are $n+1$ mutually unbiased bases. 
\end{theorem}  
 
For a proof of this result see \cite{bbrv02,ACW:04,KT:94}. It was 
shown in \cite{ACW:04} that this construction is in a certain sense 
special. This was already shown in \cite{KT:94} in the language of 
ODs. (See latter part of the proof of Theorem \ref{th:irreducible} of 
Section \ref{irreducibility}). 
 
\begin{theorem}[Limitations of nice MUBs] 
\label{th:pawelnice} 
Let ${\cal N}$ be a nice error basis in dimension $n$. Then we can 
obtain at most  
\begin{equation}\label{eq:niceBound} 
\min_{p\in\pi(n)} d_p + 1 
\end{equation} 
maximal commuting classes according to Lemma~\ref{lemma:constr_MUB} by 
partitioning ${\cal N}$, where $\pi(n)$ denotes the primes dividing 
$n$ and $d_p$ denotes the largest power of $p$ dividing $n$. 
 
Moreover, a nice error basis can be partitioned according to 
Lemma~\ref{lemma:constr_MUB} into $n+1$ commuting classes iff $n$ is a 
prime power and moreover the nice error basis is equivalent to the 
unitary error basis in eq.~(\ref{eq:tensorGenPauli}). 
\end{theorem} 
  
The idea behind this proof\cite{ACW:04} is that the maximal commuting classes 
correspond to certain collections of abelian subgroups of the index 
group $G$ of order $n$. Using some group-theoretic arguments it can be 
shown that the sizes of such collections are bounded by the smallest 
prime power contained in $n$ plus one. 

\section{Monomial Error Bases and Monomial MUBs}
\label{sec:monodef}

We say that an orthonormal basis of $\C^n$ is monomial  
if the basis vectors are the common eigenvectors of a maximal  
commuting class containing only monomial matrices. Recall that an  
$n\times n$-matrix is called monomial if each row and each column of  
it has exactly one nonzero entry.   
  
Now we can define what we mean for a collection  
$\pi=\{\cB_1,\ldots,\cB_\mu\}$ of mutually unbiased bases to be  
monomial.  
 
\begin{definition}[Monomial MUBs] 
\label{def:monomubs}
Let $\pi=\{\cB_1,\ldots,\cB_\mu\}$ be a collection of mutually unbiased 
bases. We say that $\pi$ is monomial if there is a unitary matrix $U$ 
such that the maximal commuting classes $\cC_1,\ldots,\cC_\mu$ from 
Lemma~\ref{lemma:constr_MUB} corresponding to the $U$-conjugate 
collection $U^\dagger\pi U:=\{U^\dagger \cB_1 U,\ldots, U^\dagger 
\cB_\mu U\}$ contain only monomial matrices. 
\end{definition} 

\begin{lemma}
\label{lem:monoabelian}
The standardized Hadamards $H_i^{\dagger} H_j$ generated by a set of
MUBs $H_1,\ldots,H_m$ are character tables of abelian groups iff the
MUBs are monomial\cite{Oscar}.
\end{lemma}

One application of known results on ODs to MUBs that we will  
see in Section \ref{monomial} concerns {\em monomial} matrices and MUBs.  
  
\begin{observation} 
\label{ob:nicemonocomplete} 
Notice that the unitary error basis in eq.~(\ref{eq:tensorGenPauli}) 
consists of monomial matrices and hence the complete MUB construction 
of Theorem \ref{th:oscarcomplete} yields monomial MUBs, as will any 
construction based on nice error bases, by Theorem \ref{th:pawelnice}. 
\end{observation}

For prime power dimensions $n=p^e$, there is only one other 
construction that attains the upper bound of $n+1$ MUBs, besides the 
construction in Theorem~\ref{th:oscarcomplete}. This construction is 
based on exponential sums in finite fields and Galois rings that 
attains the upper bound of $n+1$ MUBs \cite{KR:03}. 

There are 3 cases in this construction. 
A manuscript of \cite{Aschbacher} shows that at least 2 of these
cases are monomial and strongly indicates this for the 3rd case as well.
This is done by showing that the corresponding MUBs can be obtained 
equivalently (after a basis change) 
by partitioning the nice error bases in eq.~(\ref{eq:tensorGenPauli}). 
We get the following strong conjecture: 
 
\begin{conjecture}
\label{con:nicemonocomplete}
The complete collection of MUBs obtained by \cite{KR:03} 
is also monomial and obtained by partitioning nice error bases.  
This would imply that all known complete collections of MUBs
have these two properties.
\end{conjecture}

What about the single known construction of 
{\it Latin} MUBs \cite{Wocjan}  which are 
non-complete collections of MUBs? 
Are they monomial (by Definition \ref{def:monomubs})? 

Let us recall some facts about these Latin MUB collections. They exit
only when the dimension $n$ is a square and have at most $\sqrt{n}+1$
MUBs.  Each MUB in the collection (represented as a $n\times n$
generalized Hadamard as in Section \ref{intro}) is obtained using an
embedding operation (a type of tensoring) of 2 ingredient matrices:
(a) the $n \times \sqrt{n}$ incidence matrix of one of the parallel
classes of a net obtained from mutually orthogonal Latin squares of
dimension $\sqrt{n}$; and (b) a $\sqrt{n} \times \sqrt{n}$ generalized
Hadamard matrix.

The following theorem   shows that 
monomial Latin MUB collections exist.
 
\begin{observation} 
\label{ob:monopartial} 
If the $\sqrt{n}\times \sqrt{n}$ generalized Hadamard 
matrix used in the construction of a Latin MUB  collection $C$
is the character table of any finite abelian group of 
order $\sqrt{n}$,
then  $C$ is a monomial MUB collection.
%POB: I am no longer sure about this:
%and vice-versa up to a trivial equivalence class.
\end{observation} 

\begin{proof}
(Sketch).  This proof follows from the details of the Latin MUB
construction \cite{Wocjan} and Lemma \ref{lem:monoabelian}.  If the
$(i,j)^{th}$ element of the $\sqrt{n}\times \sqrt{n}$ generalized
Hadamard is $h_{i,j}$, and if this Hadamard matrix is a character
table, then $h_{i,j}h_{i,k} = h_{i,j+k}$.  The standardized Hadamard
matrices $H_i^{\dagger}H_j$ we get from the Latin MUB collection
$H_1,\ldots,H_m$ will be some permutations applied to
$H^{\dagger}\otimes H$.  Clearly the tensoring operation will not
destroy the abelian group character table structure.  The standard
construction uses the DFT Hadamard, which is the character table of
addition and gives an example in all dimensions.

% !!!!!!!!
% Each element of the latin hadamard is h_{(i|j), (k|l)}
% if it is character table, then
% h_{ (i|j), (k|l) } h_{(i|j), (m|n) } = h_{ (i|j), (k|l) + (m|n) }
% if (k|l) + (m|n) = (k+m|l+n) we are definitely in business, but
% it is not clear to me that is the only possible case.  Couldn't
% it be true that (k|l) + (m|n) is not reducible to (k+m|l+n)?
% If so, the underlying Hadamard might not be a character table.
% I don't think this matters much since the Latin MUB construction
% builds bigger MUBs out of smaller Hadamards, and not the other
% way around, but it would be nice to answer the question if possible.

\end{proof}

%!!!!!!!!!!!!!!!!!!!!!!!!!!!!!!!!!!
%the generalized Hadamard matrix is the character table of a finite 
%abelian group then the answer is yes. However, if the generalized 
%Hadamard matrix is any Hadamard matrix, then I don't know. 
% 
%{\bf A necessary and sufficient condition for that is:} 
% 
%Let $H$ be any generalized Hadamard matrix of order $s$. Then there is 
%a collection of $s$ monomial commuting matrices that are orthogonal 
%with respect to the trace inner product such that their common 
%eigenvectors are the column vectors of the Hadamard matrix. 
% 
%If $H$ is the character table of the cyclic group of oder $n$ (i.e., 
%the DFT matrix), then we can take the monomial matrices 
%$1,S,\ldots,S^{n-1}$, where $S$ is the cyclic shift. If $H$ is the 
%character table of any abelian group, the the commuting monomial 
%matrices are tensor products of shift operators. 
% 
%I don't know if this is true for arbitrary Hadamard matrices. I don't 
%think so. 
% 
%!!!!!!!!!!!!!!!!!!!!!!!!!!!!!!!!!!!!!!!!!!!!!!!!!!!!!!!!!!!!!!!!!!!!!!!! 

\iffalse
The following strong conjecture indicated by \cite{Oscar}
shows that the converse is also true.

\begin{conjecture}
The converse of Observation \ref{ob:monopartial} holds. 
\end{conjecture}
\fi

The following observation investigates to what extent the 
limitations of (partial) MUB collections obtained from nice error bases
(given by Theorem \ref{th:pawelnice})
affect (partial) MUB collections, in general.
Specifically, we can ask 
whether any (partial) collection of maximal commuting classes of 
orthogonal unitaries can be extended into a nice error basis.
By Observation \ref{ob:monopartial} this question is easily
answered in the negative for 
(partial) {\sl monomial} MUB collections. 

\begin{observation}
\label{ob:mononice} 
The Latin MUBs' give a collection - of 
maximal commuting classes of 
monomial matrices -  which cannot be extended into 
a nice error basis.
I.e., there is no nice error basis that includes
all the matrices in this collection.
\end{observation}

\begin{proof}
By Observation \ref{ob:monopartial}, choosing Hadamard matrices
derived from character tables, the Latin MUB construction of
\cite{Wocjan} yields a partial collection $\pi$ of monomial MUBs.  As
shown in \cite{Wocjan}, there are infinitely many dimensions where
sufficiently many mutually orthogonal latin squares exist whereby the
Latin MUB collection is large enough to beat the
reduce-to-minimum-prime-power lower bound.  Theorem \ref{th:pawelnice}
then implies that $\pi$ cannot be extended into a nice error basis.
\end{proof}

\section{Lie algebras and orthogonal decompositions}  
\label{liebasic}  
In this section we introduce orthogonal decompositions (ODs) of  
complex simple Lie algebras following \cite{KT:94}.  
  
We refer the reader to \cite{Graaf:00} for an introduction to Lie  
algebras, which have numerous applications. 
Some of the notions will be explained with a simple example.  
  
Let $\cL$ be a Lie algebra. A Cartan subalgebra of $\cL$ it is a  
maximal subalgebra $\cH$ that is self-normalizing, i.e., if  
$[g,h]\in\cH$ for all $h\in\cH$, then $g\in\cH$ as well.  If $\cL$ is  
simple\footnote{The results of this manuscript only concern the simple  
Lie algebra $sl_n(\C)$ consisting of traceless complex matrices of  
size $n\times n$.}, then all Cartan subalgebras of $\cL$ are abelian.  
  
As a vector space, every complex simple Lie algebra $\cL$ can be  
decomposed into a direct sum of Cartan subalgebras $\cH_i$  
\begin{equation}\label{eq:directSum}  
\cL = \cH_0 \oplus \cH_1 \oplus \cdots \cH_h\,,  
\end{equation}  
where $h$ is the Coxeter number \cite[page 12]{KT:94}. The Killing  
form  
\[  
K(A,B)=\tr(\ad A \cdot \ad B)  
\]  
is non-degenerate on $\cL$; here $\ad A$ denotes a linear operator on  
$\cL$ mapping $C\in\cL$ to $[A,C]$. The same holds for the restriction  
of the Killing form to any Cartan subalgebra $\cH$. If these two facts  
are combined by requiring all components of the decomposition in  
(\ref{eq:directSum}) to be pairwise orthogonal with respect to the   
Killing form $K$  
\begin{equation}  
K(\cH_i,\cH_j)=0\quad\mbox{for $i\neq j$,}  
\end{equation}  
then we obtain a so-called {\em orthogonal decomposition (OD)} of $\cL$  
\cite{KT:94}.  
\begin{example}  
A simple example is the special linear Lie algebra  
$\cL:=sl_2(\C)$. It has three basis vectors $\{X,Y,Z\}$, where $[X,Y]=2Z$:  
\[  
X:=\left(  
\begin{array}{rr}  
0 & 1 \\  
1 & 0  
\end{array}  
\right)\,,\quad  
Y:=\left(  
\begin{array}{rr}  
0 & -1 \\  
1 & 0  
\end{array}  
\right)\,,\quad  
Z:=\left(  
\begin{array}{rr}  
1 & 0 \\  
0 & -1  
\end{array}  
\right)\,.  
\]  
The other brackets are given by $[X,Z]=2Y$ and $[Y,Z]=2X$. In the  
adjoint representation, with the ordered basis $\{X,Y,Z\}$, these  
elements are represented by  
\begin{equation}  
\ad X=  
\left(  
\begin{array}{rrr}  
0 & 0 & 0\\  
0 & 0 & 2\\  
0 & 2 & 0  
\end{array}  
\right)\,,\quad  
\ad Y=  
\left(  
\begin{array}{rrr}  
 0 & 0 & 2\\  
 0 & 0 & 0\\  
-2 & 0 & 0  
\end{array}  
\right)\,,\quad  
\ad Z=  
\left(  
\begin{array}{rrr}  
0  & -2 & 0\\  
-2 &  0 & 0\\  
0  &  0 & 0  
\end{array}  
\right)\,.  
\end{equation}  
In this case, the Killing form can also be written as   
$K(u,v):=u^T \cK v$, where  
\begin{equation}  
\cK=\left(  
\begin{array}{rrr}  
8 &  0 & 0 \\  
0 & -8 & 0 \\  
0 &  0 & 8   
\end{array}  
\right)  
\end{equation}  
From this it is clear that   
\begin{equation}  
\cL=\langle X \rangle_\C \oplus  
\langle Y \rangle_\C \oplus  
\langle Z \rangle_\C   
\end{equation}  
is an OD.  
\end{example}

%%%XXXXXXXXXXXXXXXXXXXXXXMotivations for studying OD'sXXXXXXXXXXXXXXXXX

In the following, we work only with the special linear Lie algebra  
$sl_n(\K)$ (consisting of traceless matrices of size $n\times n$)  
because it is related to the MUB problem. This Lie algebra is  
simple and has so-called type $A_{n-1}$. 
We have already addressed the general motivation for studying orthogonal 
decompositions of complex simple Lie algebras. In the case of the 
special linear Lie algebra $sl_n(\C)$, there is in addition   
a natural link between orthogonal decompositions of Lie algebras
and orthogonal pairs of maximal abelian subalgebras in von Neumann algebras
which are investigated by P. de la Harpe, V. F. R. Jones,
S. Popa, U. Haagerup, and others 
\cite{HJ}, \cite{P}, \cite{KT:94}, Section 1.5. 

The Killing form of two matrices $A$ and $B$ in $sl_n(\K)$ can  
be expressed as the trace of their product. More precisely, we have:  
  
\begin{lemma}\label{lem:Killing}  
Let $A,B\in sl_n(\K)$. Then the Killing form $K(A,B):=\tr(\ad A \cdot  
\ad B)$ is equal to  
\begin{equation}  
2n \tr(A B)\,,  
\end{equation}  
where $\tr(A B)$ denotes the trace of the product of matrices $A$ and  
$B$.   
\end{lemma}

\section{Establishing the Connection between ODs and MUBs}  
\label{connection}  
We now have the background necessary to show that MUBs give rise to   
orthogonal Cartan subalgebras and viceversa when these Cartan subalgebras 
are closed under the adjoint operation $\dag$. 
  
\begin{theorem}  
\label{th:MUBstoOD}  
If there are $\mu$ pairwise mutually unbiased bases  
$\cB_1,\cB_2,\ldots,\cB_\mu$ of $\K^n$ then there are $\mu$ Cartan  
subalgebras $\cH_1,\cH_2,\ldots,\cH_\mu$ of $sl_n(\K)$ that are  
pairwise orthogonal with respect to the Killing form, where $\K=\R$ or  
$\K=\C$. In particular, a complete collection of MUBs in $\C^n$ gives rise to  
an orthogonal decomposition of $sl_n(\C)$.  
\end{theorem}  
  
\begin{proof}  
Let $\cB=\{\ket{\psi_1},\ket{\psi_2},\ldots,\ket{\psi_n}\}$ be  
an orthonormal basis of $\K^n$. Define $\cH$ to be the linear subspace  
of $sl_n(\K)$ consisting of all traceless matrices that are diagonal  
in the basis $\cB$. Clearly, $\cH$ is abelian, i.e., all matrices in  
$\cH$ commute. Furthermore, $\cH$ is maximal, i.e., $\cH$ cannot be  
enlarged and remain abelian at the same time. Therefore, it only  
remains to show that $\cH$ is self-normalizing in order to prove that  
it is a Cartan subalgebra.  Let $A\in sl_n(\K)$. If $[A,D]\in \cH$ for  
all $D\in\cH$, then this implies that $A\in\cH$. This can be checked  
by using the matrices  
$D_{ij}:=\ket{\psi_i}\bra{\psi_i}-\ket{\psi_j}\bra{\psi_j}$ and  
looking at the form of the resulting commutators  
$[A,D_{ij}]$. Consequently, $\cH$ is a Cartan subalgebra.  
  
Now we prove that two mutually unbiased bases give rise to two Cartan  
subalgebras that are orthogonal with respect to the Killing form. Let  
$\cB':=\{\ket{\phi_1},\ket{\phi_2},\ldots,\ket{\phi_d}\}$ be an  
orthonormal basis of $\K^n$ such that $\cB$ and $\cB'$ are mutually  
unbiased. Define $\cH'$ to be the Cartan subalgebra consisting of  
traceless matrices that are diagonal with respect to $\cB'$. Let  
$A\in\cH$ and $B\in\cH'$ be two arbitrary matrices in $\cH$ and  
$\cH'$, respectively. They can be expressed as  
\[  
A:=\sum_{i=1}^n a_i \ket{\psi_i}\bra{\psi_i}\quad\mbox{and}\quad  
B:=\sum_{j=1}^n b_j \ket{\phi_j}\bra{\phi_j}\,,  
\]  
where $\sum_i a_i=\sum_i b_i=0$. Now it is easily seen that $\cH$ and  
$\cH'$ are orthogonal with respect to the Killing form:  
\begin{eqnarray*}  
K(A,B) & = & 2n \tr(AB) \\  
& = &  
\sum_i a_i \sum_j b_j   
\langle \phi_j | \psi_i \rangle  
\langle \psi_i | \phi_i \rangle \\  
& = &  
\sum_i a_i \sum_j b_j \frac{1}{d} \\  
& = & 0  
\end{eqnarray*}   
The first equality follows from Lemma~\ref{lem:Killing} and the second  
from the fact that $\cB$ and $\cB'$ are mutually unbiased. This  
completes the proof.  
\qed  
\end{proof}  
  
Theorem \ref{th:MUBstoOD} shows that a (complete) collection of
mutually unbiased bases always gives rise to (a decomposition into)
orthogonal Cartan subalgebras of $sl_n(\C)$. In the following theorem,
we show when the converse holds: we characterize orthogonal Cartan
subalgebras of $sl_n(\C)$ that give rise to to mutually unbiased
bases. Let $\dagger$ denote the adjoint operation, i.e., the
involutory map $A\mapsto A^\dagger:=\bar{A}^t$, where $\bar{C}$ means
that we take the complex conjugate of each entry of a matrix $C$ and
$C^t$ means that we transpose $C$.
  
\begin{theorem}  
\label{th:dagger}  
Collections of $\mu$ Cartan subalgebras $\cH_1,\ldots,\cH_\mu$ of  
$sl_n(\C)$ that are pairwise orthogonal with respect to the Killing  
form and are closed under the adjoint operation, i.e., the   
involutory map $\dagger$, correspond to  
collections of $\mu$ mutually unbiased bases $\cB_1,\ldots,\cB_\mu$ and  
vice versa.  
\end{theorem}  
  
\begin{proof}  
The Cartan subalgebras constructed from the orthonormal bases  
in Theorem~\ref{th:MUBstoOD} are clearly $\dagger$-closed.  
  
Let $\cH$ be a Cartan subalgebra that is $\dagger$-closed. Then all  
matrices in $\cH$ are normal.\footnote{Recall that an operator $A$ is  
normal if and only if it commutes with its adjoint $A^\dagger$, i.e.,  
$AA^\dagger=A^\dagger A$. Normal matrices are precisely those which  
can be diagonalized.}   
  
%We don't need this: Let $B_1,\ldots,B_{d-1}$ be a basis of $\cH$ and  
%$A$ an arbitrary operator in $\cH$. Its adjoint $A^\dagger$ can be  
%expressed as a linear combination of the basis matrices  
%$B_1,\ldots,B_{d-1}$ since $\cH$ is $\dagger$-closed. This shows that  
%$A$ commutes with $A^\dagger$ because $A$ commutes with all basis  
%operators.   
  
From standard linear algebra, it follows that   
all matrices in $\cH$ can be simultaneously  
diagonalized since they are normal and they all commute. Let  
$\cB:=\{\ket{\phi},\ldots,\ket{\phi_n}\}$ be an orthonormal basis  
consisting of the common eigenvectors of the matrices in $\cH$.  
  
Let $\cH'$ be a second $\dagger$-closed Cartan subalgebra that is  
orthogonal to $\cH$ with respect to the Killing form. Denote by  
$\cB':=\{\ket{\psi},\ldots,\ket{\psi_n}\}$ the corresponding  
orthonormal basis consisting of the common eigenvectors of all  
matrices in $\cH'$. We show that $\cB$ and $\cB'$ are mutually  
unbiased.  
  
Let $S$ be the matrix with entries $s_{ij}:=|\langle\phi_i |  
\psi_j\rangle|^2$ for $i,j=1,\ldots,n$. Observe that $S$ is a double  
stochastic matrix. This follows from the fact that $\cB$ and $\cB'$  
are orthonormal bases. Assume that $\cB$ and $\cB'$ are not mutually  
unbiased. Then $S$ is not equal to $\frac{1}{n}J$ where $J$ denotes the  
all-one-matrix. Consequently, we may assume without loss of generality  
(by suitably permuting rows and columns of $S$) that $s_{11}\neq  
s_{21}$. Now there must be a column $k\neq 1$ such that  
\[  
s_{11}-s_{21}\neq s_{1k}-s_{2k}\,.  
\]  
Otherwise, we would have $\sum_j s_{1j}-s_{2j}\neq 0$ contradicting  
the fact that the rows are probability distributions. Again without  
loss of generality we may assume that $k=2$. But this shows that the  
Killing form of the matrices  
\[  
A:=\ket{\phi_1}\bra{\phi_1}-\ket{\phi_2}\bra{\phi_2}\quad\mbox{and}\quad  
B:=\ket{\psi_1}\bra{\psi_1}-\ket{\psi_2}\bra{\psi_2}  
\]  
is  
\[  
K(A,B)=2n (s_{11}-s_{21}-(s_{12}-s_{22}))  
\]   
not zero even though $A\in\cH$ and $B\in\cH'$. But this is a  
contradiction to the assumption that $\cH$ and $\cH'$ are orthogonal  
with respect to the Killing form. This completes the proof.  
\qed  
\end{proof}  
  
Next we relate Theorem \ref{th:dagger} to   
Lemma \ref{lemma:constr_MUB}, a known result about MUBs.  
  
\begin{proposition}  
\label{th:daggeroscar}  
Any $\dag$-closed Cartan subalgebra of $sl_n(\C)$ has a basis  
of unitary matrices that are pairwise orthogonal with respect to  
the Killing form. Viceversa, any maximal commuting class of   
orthogonal unitary matrices spans a $\dag$-closed Cartan  
subalgebra.  
Thus Theorem \ref{th:dagger} is equivalent to  
Lemma \ref{lemma:constr_MUB}.  
\end{proposition}  
  
Finally, we state a 20 year old conjecture of \cite{KT:94}  
which implies a conjecture that has existed in the   
MUBs community.   
  
\begin{conjecture}  
\cite{KT:94}  
The Lie algebra $\cL = sl_{n}(\C)$ has an OD only if $n$ is a prime power.  
\end{conjecture}  
  
This implies, using Theorem \ref{th:MUBstoOD} that   
  
\begin{conjecture}  
A complete collection of $n+1$ MUBs exists only in prime power dimensions $n$.  
\end{conjecture}  
  
OD constructions   
are known (cf. the {\it standard} OD in \cite{KT:94}, Chap 1.) analogous to the   
optimal MUB constructions    
as in Theorem \ref{th:oscarcomplete}  
that meet the $n+1$   
upper bound for prime power dimensions.   
They have much the same  
properties such as   
niceness and monomiality (as in Observation \ref{ob:nicemonocomplete}).   
This will be discussed further in   
Section \ref{irreducibility}, where   
we will see  
that in fact the two constructions are essentially equivalent  and satisfy  
further strong requirements.

\subsection{A new upper bound for a common type of MUB}  
\label{monomial}  
Both conjectures of the previous section remain open even in the   
smallest dimension $n = 6$. However, the following is shown in   
\cite{KT:94} (Chap. 1).   
A Cartan   
subalgebra $\cH$ is called {\it monomial} if it has a basis consisting of   
only monomial   
matrices.

\begin{theorem}  
\label{th:monomialcartan}  
The Lie algebra   
$sl_{6}(\C)$   
cannot have more than $3$ pairwise orthogonal   
monomial Cartan subalgebras.   
\end{theorem}  
  
This gives the first nontrivial MUB upper bound below for non-prime-power    
dimensions for  monomial MUBs.  
These are a common and natural type of MUB, since   
Observation \ref{ob:nicemonocomplete},  
and \ref{ob:monopartial} 
indicate that known constructions of MUBs 
- both complete and partial collections -  
correspond to monomial MUBs.   
  
\begin{theorem}  
\label{th:monomialmubs}  
There are no more than 3 monomial MUBs in dimension $n=6$.  
\end{theorem}  
  
The proof follows directly from Theorem \ref{th:monomialcartan},   
using the  definitions of monomial MUBs and Cartan  
subalgebras and  Theorem  
\ref{th:daggeroscar}.  
  
\section{Irreducible ODs and new properties of known MUB constructions}  
\label{irreducibility}  
  
Let $\Aut(\cL)$ be the group of automorphisms of the Lie algebra $\cL
= sl_{n}(\C)$. This group acts on $\cL$ by conjugation. We say that
two elements (resp. subsets, collections of subsets) of $\cL$ are {\it
equivalent} if one is an $\Aut(\cL)$-conjugate of the other.
Likewise, we say that an element (resp. subset, collection of subsets)
satisfying a property $\cP$ is {\it unique} if all elements
(resp. subsets, collections of subsets) satisfying $\cP$ are in the
same $\Aut(\cL)$-conjugacy class.
  
\begin{theorem}  
\label{th:standard}  
If $n$ is a prime power, then $sl_n(\C)$ admits the so-called {\it
standard} OD (cf. \cite{KT:94}, Chap. 1).  This OD consists of
monomial, $\dagger$-closed Cartan subalgebras.  By Theorem
\ref{th:daggeroscar}, it corresponds to a partition of a nice error
basis and is equivalent to the OD obtained the construction of Theorem
\ref{th:oscarcomplete}.
\end{theorem}  
  
The description of the standard OD  gives another view of the   
construction of Theorem \ref{th:oscarcomplete}   
as arising from the projective representation of degree   
$p^{m}$ of the elementary abelian group of order $p^{2m}$, using  
the desarguesian plane of order $p^{m}$.  
  
Let $\cD~:~\cL = \oplus^{n+1}_{i=1}\cH_{i}$ be an   
OD of $\cL$.  
Further, let $\Aut(\cD)$ be the subgroup of all   
automorphisms $\varphi$ of $\cL$ that preserve $\cD$   
(that is, for any $i$ there is  
$i'$ such that $\varphi(\cH_{i}) = \cH_{i'}$).   
Clearly, $\Aut(\cD)$ also acts on the space  
$\cL$ by conjugation.   
We say that  an OD $\cD$ is {\it irreducible}   
if   
$\Aut(\cD)$ acts {\it irreducibly} on $\cL$. I.e., the orbit of any   
element of $sl_n(\C)$ under this action spans all of   
$sl_n(\C)$.   
  
One of the main results of \cite{KT:94} is that the standard OD   
of Theorem \ref{th:standard} is   
essentially the unique irreducible OD.   
  
\begin{theorem}  
\label{th:irreducible}  
If Lie algebra $\cL = sl_n(\C)$ admits an irreducible OD   
$\cD$, then   
$n = p^{m}$ for some prime $p$, and $\cD$ is equivalent  
the aforementioned standard OD, except for $n = 27$.  
\end{theorem}   
  
In the case of $n = 27$, an additional irreducible OD    
$\cD$ can be constructed   
using the projective representation of degree $27$   
of the elementary abelian group   
of order $3^{6}$, together with the so-called Hering plane of order $27$.  
  
\begin{proof}  
(Sketch).  The proof in \cite{KT:94} (Chapters 4 and 5) is quite
involved and uses the classification of finite simple groups. The
first part shows that in order for the OD $\cD$ to be irreducible, the
group $\Aut(\cD)$ must have a nontrivial subgroup $S$ that fixes each
Cartan subalgebra in $\cD$: i.e., the kernel of the homomorphism
$\phi~:~\Aut(\cD) \to S_{n+1}$ is non-empty; $S_{n+1}$ is the
symmetric group of permutations, in this case, of the $n+1$ Cartan
subalgebras in the OD.  The latter part uses the fact that this
subgroup $S$ is a $p$-group and shows that the OD must be the standard
OD of Theorem \ref{th:standard}.  This part of the proof is
effectively the proof in \cite{ACW:04} that a complete partition of a
nice error bases only exists for prime power dimensions and is
equivalent to the construction in Theorem \ref{th:oscarcomplete}.
\qed
\end{proof}  
  
Next we define the irreducibility condition directly for complete MUB
collections, which intuitively indicates that the MUB collection has a
large group of symmetries.  We then prove a corresponding result to
Theorem \ref{th:irreducible} directly for MUBs, showing that the
standard MUB collection of Theorem \ref{th:oscarcomplete} is in fact
irreducible,
  
\begin{theorem}  
\label{th:irreduciblemubs}  
Let $\pi := \{\cB_1,\cB_2,\ldots,\cB_{n+1}\}$ be a complete collection
of pairwise mutually unbiased bases of $\C^n$. Consider the subgroup
$\Aut(\pi)$ consisting of all unitary matrices $X \in U_{n}(\C)$ that
preserve $\pi$, that is, for any $i$ there is $i'$ such that the sets
$X\cB_{i}X^{-1}$ and $\cB_{i'}$ are equal. Assume that $\Aut(\pi)$
acts (via conjugation) on $\cL = sl_{n}(\C)$ irreducibly.  We call
such a $\pi$ an {\it irreducible MUB collection}.  Then $n = p^{m}$ is
a prime power. Moreover, if $n \neq 27$ then such a $\pi$ is unique up
to $U_{n}(\C)$-conjugacy and the order of the members of $\pi$: it is
$U_{n}(\C)$-conjugate to the MUB collection of Theorem
\ref{th:oscarcomplete}.
\end{theorem}  
  
\begin{proof}  
Consider the OD $\cD$ corresponding to $\pi$ by Theorem   
\ref{th:MUBstoOD}. It is   
straightforward to check that $\Aut(\pi)$ acts on $\cD$. This action induces   
a homomorphism $\phi~:~\Aut(\pi) \to \Aut(\cD)$ with kernel equal  
$\Aut(\pi) \cap Z(U_{n}(\C))$. It follows by the assumption   
that $\Aut(\cD)$ acts   
irreducibly on $\cL$. By Theorem \ref{th:irreducible},  
$n = p^{m}$ is   
a prime power, and if $n\neq 27$, then $\cD$ is unique and equivalent  
to the standard OD of Theorem \ref{th:standard}.   
Observe that $\pi$ can be reconstructed uniquely from $\cD$:  each   
Cartan subalgebra $\cH_{i}$ that lies in $\cD$ can be diagonalized to yield   
a unique  
basis $\cB_{i}$ of $\C^{n}$, and $\pi$ is just the collection   
of all $\cB_{i}$'s, which, by Theorem \ref{th:standard}, is   
$U_{n}(\C)$-conjugate to the MUB collection of Theorem \ref{th:oscarcomplete}.  
\qed  
\end{proof}

Finally, we use a known result of \cite{KT:94} for small dimensions $n \le 5$  
to show that complete collections of MUBs for these dimensions are  
essentially unique, monomial, irreducible and    
obtained by partitioning a nice error basis. In fact, they    
are equivalent to the standard complete MUB collection of   
Theorem \ref{th:oscarcomplete}  
for prime power dimensions.   
  
\begin{theorem}  
\label{th:smallOD}  
The Lie Algebra $sl_{n}(\C)$   
for $n \leq 5$ has a unique, irreducible OD  equivalent to the standard OD.  
\end{theorem}  
  
We can state this result directly for MUBs as follows.  
  
\begin{theorem}  
\label{th:smallMUB}  
Let $\pi := \{\cB_1,\cB_2,\ldots,\cB_{n+1}\}$ be a complete set of   
pairwise mutually unbiased bases of $\C^n$ for $n \leq 5$.   
Then such a $\pi$ is   
unique up to $U_{n}(\C)$-conjugacy and the order of the members of $\pi$.  
which, by Theorem \ref{th:standard}, is   
$U_{n}(\C)$-conjugate to the MUB collection of Theorem \ref{th:oscarcomplete}.  
\end{theorem}  
  
\begin{proof}  
Use the connection between ODs and MUBs as in the proof of Theorem   
\ref{th:irreduciblemubs},   
and apply   
Theorem \ref{th:smallOD}.  
\qed  
\end{proof}

\section{Potential Extensions and Related Questions} 
 
We have shown that if an OD is  
irreducible, then it exists only in prime power dimensions and is  
essentially unique: monomial, $\dagger$-closed hence  
gives rise to a complete collection of MUBs equivalent to the  
the known construction of MUBs, obtained by partitioning nice error bases.  
For small prime powers $n \le 5$, ODs  
and the corresponding complete MUB collections are unique. 
For $n =6$ at most 3 monomial MUBs or Cartan subalgebras exist. 
 
To leverage the above results,  
here are some natural questions to ask.  
 
Although tempting, it is not promising to conjecture that if there is
any $\dag$-closed OD, then there is one that is irreducible.
Intuitively, irreducibility means that the OD has a large group of
symmetries and there are many instances where such statements are
known to be false, i.e., the optimal or extremal constructions often
do not exhibit significant symmetry.
 
\smallskip\noindent {\sl Question 1.}  Can one generalize the
irreducibility condition or its consequences either by (a) weakening
the condition - instead of requiring that the action of $\Aut(\cD)$ on
$sl_n(\C)$ be irreducible, require that it generates a large
irreducible component of $sl_n(\C)$; or by (b) extending the
irreducibility definition to (the group of automorphisms that
preserve) incomplete or partial collections of Cartan subalgebras.
 
On first inspection, both do not appear to be  
promising approaches.  
 
\smallskip\noindent 
{\sl Question 2.} 
Can one show that if there is any collection of Cartan subalgebras 
or MUBs or maximal commuting classes of orthogonal unitary matrices,  
then there is always a monomial collection of the same size? 
 
More general related questions are the following. 
 
\smallskip\noindent 
{\sl Question 3.} 
How do these properties (irreducibility, niceness, monomiality)  
relate in general? 
 
Some care is required in posing this type of question since these  
properties apply to different entities. 
Irreducibility is a property so far defined and studied only for complete ODs,  
(although as mentioned 
above, the definition could potentially be extended to partial collections 
of Cartan subalgebras). 
Niceness is a property of  
a complete,  
unitary error basis without the requirement of the existence of an OD.  
Monomiality is a property that is well-defined for  
any (partial) collection of  maximal commuting classes, Cartan subalgebras 
or MUBs. 
 
Results in \cite{KT:94} show that monomiality of an OD does not  
imply irreducibility of the OD. 
 
Furthermore, when refering just  to unitary error bases (without imposing  
the requirement that they be partitionable into a complete collection),  
these monomiality and niceness  
do not imply each other  
as shown by \cite{KR:03b}: 
not all nice error bases are monomial nor viceversa. 
 
In this context,  natural questions related to the results of \cite{KR:03b}  
is the following.  
 
\smallskip\noindent 
{\sl Question 4.} 
If we restrict ourselves to unitary error 
bases that give a complete collection of maximal 
commuting classes and MUBs,  
then we know that niceness of the unitary error basis imply monomiality 
by Theorem \ref{th:pawelnice} (or Theorem \ref{th:irreducible}). 
But under this condition, 
does monomiality imply niceness?
A positive answer to this question would imply that monomial complete
collections of MUBs exist only in prime power dimensions.

\medskip\noindent
{\sl Question 5.} 
Since monomiality is well-defined for any (partial) collection of 
maximal commuting classes, 
we can ask if such a collection can always be extended into 
an entire error basis that is   monomial. 
Note that if the answer to Question 4 is yes, then the answer
to Question 5 must be no, and viceversa, 
because of Observation \ref{ob:mononice}.

\section{Acknowledgements}
The work of MS supported in part by NSF Grants  EIA 02-18435, CCF 04-04116.
The work of PHT is supported in part by NSA grant H98230-04-0066.
The work of PW is supported by NSF Grant EIA 00-86038.

\bibliographystyle{plain} 
\bibliography{mub} 

\end{document}